\catcode`\@=11
\newif\if@fewtab\@fewtabtrue

{\count255=\time\divide\count255 by 60
\xdef\hourmin{\number\count255}
\multiply\count255 by-60\advance\count255 by\time
\xdef\hourmin{\hourmin:\ifnum\count255<10 0\fi\the\count255}}
\def\ps@draft{\let\@mkboth\@gobbletwo
    \def\@oddhead{}
    \def\@oddfoot
       {\hbox to 7 cm{$\scriptstyle Draft\ version:\ \draftdate$
       \hfil}\hskip -7cm\hfil\rm\thepage \hfil}
    \def\@evenhead{}\let\@evenfoot\@oddfoot}

\def\ceqno{\global\@fewtabfalse
    \ifcase\@eqcnt \def\@tempa{& & &}\or \def\@tempa{& &}
      \or \def\@tempa{&}
      \or\def\@tempa{}\fi\@tempa
{\rm(\theequation)}}

\def\aeqno#1{\global\@fewtabfalse
    \ifcase\@eqcnt \def\@tempa{& & &}\or \def\@tempa{& &}
      \or \def\@tempa{&}
      \or\def\@tempa{}\fi\@tempa
{\rm(\theequation,#1)}}

\def\label#1{\ifnum\draftcontrol=1
 \global\def\draftnote{$\scriptstyle #1$}\fi
 \@bsphack\if@filesw {\let\thepage\relax
   \def\protect{\noexpand\noexpand\noexpand}%
\xdef\@gtempa{\write\@auxout{\string
      \newlabel{#1}{{\@currentlabel}{\thepage}}}}}\@gtempa
   \if@nobreak \ifvmode\nobreak\fi\fi\fi
  \@esphack}

\def\alabel#1#2{\label{#1}\global\@fewtabfalse
    \ifcase\@eqcnt \def\@tempa{& & &}\or \def\@tempa{& &}
      \or \def\@tempa{&}
      \or\def\@tempa{}\fi\@tempa
{\hbox to 3cm{\phantom{\rm(\theequation,#2)}
\draftnote \hfil}\hskip -3cm {\rm(\theequation,#2)}}}

\def\clabel#1{\label{#1}\global\@fewtabfalse
    \ifcase\@eqcnt \def\@tempa{& & &}\or \def\@tempa{& &}
      \or \def\@tempa{&}
      \or\def\@tempa{}\fi\@tempa
{\hbox to 3cm{\phantom{\rm(\theequation)}
\draftnote \hfil}\hskip -3cm{\rm(\theequation)}}}

\def\eqnarray{\def\draftnote{{}}\global\@fewtabtrue
\stepcounter{equation}\let\@currentlabel=\theequation
\global\@eqnswtrue
\global\@eqcnt\z@\tabskip\@centering\let\\=\@eqncr
$$\halign to \displaywidth\bgroup\@eqnsel\hskip\@centering\@eqcnt\z@
  $\displaystyle\tabskip\z@{##}$&\global\@eqcnt\@ne
  \hskip 1\arraycolsep \hfil${##}$\hfil
  &\global\@eqcnt\tw@ \hskip 1\arraycolsep
$\displaystyle\tabskip\z@{##}$
\hfil  \tabskip\@centering&\global\@eqcnt\thr@@\llap{##}\tabskip\z@
\cr}

\def\endeqnarray{\@@eqncr\egroup
      \global\advance\c@equation\m@ne$$\global\@ignoretrue}

\def\@eqnnum{\hbox to 3cm{\phantom{\rm(\theequation)} \draftnote
			 \hfil}\hskip -3cm {\rm(\theequation)}}

\def\@@eqncr{\let\@tempa\relax
    \ifcase\@eqcnt \def\@tempa{& & &}\or \def\@tempa{& &}
      \or \def\@tempa{&}
      \or\def\@tempa{}
\fi\@tempa
\if@eqnsw
\if@fewtab\@eqnnum\fi
\stepcounter{equation}\fi\global
\@eqnswtrue\global\@eqcnt\z@\global\@fewtabtrue\cr}

\def\draftcite#1{\ifnum\draftcontrol=1#1\else{}\fi}

\def\@lbibitem[#1]#2{\item{}\hskip -3cm \hbox to 2cm
{\hfil$\scriptstyle\draftcite{#2}$}\hskip
1cm[\@biblabel{#1}]\if@filesw
     {\def\protect##1{\string ##1\space}\immediate
      \write\@auxout{\string\bibcite{#2}{#1}}}\fi\ignorespaces}

\def\@bibitem#1{\item\hskip -3cm \hbox to 2cm
{\hfil $\scriptstyle\draftcite{#1}$}\hskip 1cm
\if@filesw \immediate\write\@auxout
       {\string\bibcite{#1}{\the\value{\@listctr}}}\fi\ignorespaces}

\def\nsection#1{\section{#1}\setcounter{equation}{0}}

\font\tendl=msbm10  scaled \magstep1
\font\sevendl=msbm7 scaled \magstep1
\font\fivedl=msbm5 scaled \magstep1
\font\tengl=eufm10  scaled \magstep1
\font\sevengl=eufm7 scaled \magstep1
\font\fivegl=eufm5 scaled \magstep1

\newfam\dlfam \def\dl{\fam\dlfam\tendl} 
\textfont\dlfam=\tendl \scriptfont\dlfam=\sevendl
\scriptscriptfont\dlfam=\fivedl
\newfam\glfam  
\textfont\glfam=\tengl \scriptfont\glfam=\sevengl
\scriptscriptfont\glfam=\fivegl

\def\draftdate{\number\month/\number\day/\number\year\ \ \ \hourmin }

\global\def\draftcontrol{0}
\catcode`\@=12
\def\tilde{\widetilde}

\documentstyle[11pt]{article}

\renewcommand{\theequation}{\thesection.\arabic{equation}}
\def\theequation{{\thesection.\arabic{equation}}}

\newcommand{\be}{\begin{eqnarray}}
\newcommand{\en}{\end{eqnarray}\vs 0.5 cm}

\newcommand{\vs}{\vskip}

\newcommand{\NP}{{{\dl P}}}
\newcommand{\NC}{{{\dl C}}}
\newcommand{\NZ}{{{\dl Z}}}

\newcommand{\qq}{\begin{eqnarray}}
\newcommand{\de}{\bar\partial}
\newcommand{\da}{\partial}
\newcommand{\ee}{{\rm e}}

\newcommand{\qqq}{\end{eqnarray}}

\newcommand{\tr}{\hbox{tr}}

\newcommand{\CA}{{\cal A}}

\newcommand{\CC}{{\cal C}}

\newcommand{\CG}{{\cal G}}

\newcommand{\CK}{{\cal K}}
\newcommand{\CL}{{\cal L}}

\newcommand{\CN}{{\cal N}}
\newcommand{\CO}{{\cal O}}
\newcommand{\CP}{{\cal P}}

\newcommand{\CS}{{\cal S}}

\newcommand{\CW}{{\cal W}}

\newcommand{\s}{\hspace{0.05cm}}
\newcommand{\m}{\hspace{0.025cm}}

\newcommand{\imtau}{{\tau_2}}
\begin{document}

\begin{center}

{\Large \bf HITCHIN SYSTEMS AT LOW GENERA}\\

\vskip 1cm

{\sc Krzysztof Gaw\c{e}dzki}\\
I.H.E.S., C.N.R.S., F-91440 Bures-sur-Yvette, France\\

\vskip 0.5 cm

{\sc Pascal Tran-Ngoc-Bich}\\
Universit\'e de Paris-Sud, F-91405 Orsay, France

\vskip 1cm

\vskip 1cm 
{\small
\begin{center}
{\bf Abstract}
\end{center}
\begin{quote}
The paper gives a quick account of 
the simplest cases of the Hitchin integrable systems
and of the Knizhnik-Zamolodchikov-Bernard connection
at genus $0$, $1$ and $2$. In particular,
we construct the action-angle variables of the
genus $2$ Hitchin system with group $SL_2$ by exploiting 
its relation to the classical Neumann integrable systems.
\end{quote}
}
\end{center}

\vskip 1.3cm

\nsection{Hitchin systems}

\noindent As was realized by Hitchin in \cite{Hitch},
a large family of integrable systems may be obtained
by a symplectic reduction of a chiral 2-dimensional
gauge theory. Let $\Sigma$ denote a closed Riemann surface 
of genus $g$ and let $G$ be a complex Lie group which
we shall assume simple, connected and simply connected. 
We shall denote by $\CA$ the space of $Lie(G)$-valued
0,1-gauge fields\footnote{one may work in a fixed
smoothness class and use the Sobolev norms to define 
topology in $\CA$} $A=A_{\overline{z}}\,d\overline{z}$
on $\Sigma$. Hitchin's construction \cite{Hitch} associates 
to $\Sigma$ and $G$ an integrable system obtained by a symplectic 
reduction of the infinite-dimensional complex symplectic 
manifold $T^*\CA$ of pairs $(A,\Phi)$ where $\Phi=\Phi_{z}\, dz$ 
is a $Lie(G)$-valued 1,0-Higgs field. The holomorphic symplectic 
form on $T^*\CA$ is
\qq
\int_\Sigma\tr\,\, \delta\Phi\s\s\delta A
\qqq
where $\tr$ stands for the Killing form on $Lie(G)$ normalized
so that $\tr\,\phi^2=2$ for the long roots $\phi$.
The local gauge transformations $h\in{\CG}\equiv Map(\Sigma,G)$
act on $T^*\CA$ by 
\qq
A\longmapsto {}^hA\,\equiv\, hAh^{-1}+h\de h^{-1},\ \ \quad
\Phi\longmapsto {}^h\Phi\,\equiv\, h\Phi h^{-1}
\qqq
preserving the symplectic form. The corresponding
moment map $\,\mu:\, T^*\CA\longrightarrow Lie(\CG)^*\cong$
$\wedge^2(\Sigma)\otimes Lie(G)\,$ takes the form
\qq
\mu(A,\Phi)\ =\ \de\Phi+A\Phi+\Phi A.
\qqq
The symplectic reduction gives the reduced phase space
\qq
\CP\ =\ {\mu}^{-1}(\{0\})\,/\,\CG
\label{CP}
\qqq
with the symplectic structure induced from that of $T^*\CA$.
$\CP$ may be identified with the complex cotangent 
bundle $T^*\CN$ to the orbit space $\CN=\CA/\CG$
and $\CN$, in turn, with the moduli space of holomorphic 
$G$-bundles on $\Sigma$. More precisely,
care should be taken to avoid non-generic bad orbits
in order to obtain tractable orbit spaces. This may be done 
by considering only gauge fields $A$ leading to stable 
$G$-bundles forming smooth moduli space $\CN_s$
or those leading to semi-stable bundles giving rise
to a, generally singular, compactification $\CN_{ss}$
of $\CN_s$. In what follows we shall be somewhat cavalier 
about such details. 
\vskip 0.4cm

The Hitchin system has $\CP$ as its phase space.
Its Hamiltonians are obtained the following way. 
Let $p$ be a homogeneous $Ad$-invariant polynomial
on $Lie(G)$ of degree $d_p$. Then
\qq
h_p(A,\Phi)\ =\ p(\Phi)=p(\Phi_z)(dz)^{d_p}
\qqq
defines a $d_p$-differential on $\Sigma$ which
is holomorphic if $\mu(A,\Phi)=0$. Since 
$h_p$ is constant on the orbits of $\CG$,
it descends to the reduced phase space:
\qq
h_p\,:\,\CP\longrightarrow H^{0}(K^{d_{p}})\m.
\qqq
Here $K$ stands for the canonical bundle 
(of covectors $\propto dz$) and $H^0(K^{d_p})$
is the (finite-dimensional) vector space of the holomorphic
$d_p$-differentials on $\Sigma$. The (components of)
$h_p$ Poisson-commute (they Poisson-commute already 
as functions on $T^*\CA$ since they depend only 
on the "momenta" $\Phi$). The point of Hitchin's
construction is that, by taking a complete 
system of polynomials $p$, one obtains on $\CP$
a complete system of Hamiltonians in involution. 
For the matrix groups, the values of Hamiltoniens 
$h_p$ at a point of $\CP$ may be encoded 
in the spectral curve $\CC$ obtained by solving 
the characteristic equation
\qq
\det(\Phi-\xi)\,=\,0
\label{det}
\qqq
for $\xi\in K$. The spectral curve of the eigenvalues $\xi$ 
is a ramified cover of $\Sigma$. The corresponding eigenspaces 
of $\Phi$ form then a holomorphic line bundle over $\CC$
belonging to a subspace of the Jacobian of $\CC$
on which the Hamiltonians $h_p$ induce linear flows.
\vskip 0.4cm

For the quadratic polynomial $p_2={1\over2}\tr$, 
the map $h_{p_2}$ takes values in the space
of holomorphic quadratic differentials $H^0(K^2)$. This is
the space cotangent to the moduli space of complex curves $\Sigma$.
Variations of the complex structure of $\Sigma$ are described 
by Beltrami differentials $\delta\mu=\delta\mu_{\bar z}^z\,
\partial_z d\bar z$ such that $z'=z+\delta z$ with
$\da_{\bar z}\delta z=\delta\mu_{\bar z}^z$ gives new 
complex coordinates. The Beltrami differentials $\delta\mu$ may 
be paired with quadratic differentials $\beta$ by 
\qq
(\beta,\delta\mu)\ \mapsto\ \int_{\Sigma}\beta\,\delta\mu. 
\label{pair}
\qqq
The differentials $\delta\mu=\de(\delta\xi)$, where $\delta\xi$ 
is a vector field on $\Sigma$, describe variations of the complex 
structure due to diffeomorphisms of $\Sigma$ and they pair 
to zero with $\beta$. The quotient space $H^1(K^{-1})$
of differentials $\delta\mu$ modulo $\de(\delta\xi)$
is the tangent space to the moduli space of curves $\Sigma$ 
and $H^0(K^2)$ is its dual. The pairing (\ref{pair}) defines 
then for each $[\delta\mu]\in H^1(K^{-1})$
a Hamiltonian 
\qq
h_{\delta\mu}\ \equiv\ \int_\Sigma h_{p_2}\m\delta\mu\m.
\label{hmu}
\qqq
The Hamiltonians $h_{\delta\mu}$ Poisson-commute 
for different $\delta\mu$.
\vskip 0.4cm

Hitchin's construction possesses a natural generalization
\cite{Mark}\cite{Nekr}\cite{EnRub}\cite{GaFal}. Let 
$x_n\in\Sigma$ be a finite family of distinct points in $\Sigma$
and $\CO_n$ a family of (co)adjoint orbits in $Lie(G)^*\cong Lie(G)$. 
\qq
\CO\ =\ \{\,\sum\limits_n\lambda_n\delta_{x_n}\ \vert\ \lambda_x
\in{\CO}_n\,\},
\qqq
where $\delta_x$ stands for the Dirac delta measure at $x$, 
forms a coadjoint orbit of the group $\CG$ of local gauge 
transformations. In the symplectic reduction we may replace 
definition (\ref{CP}) with
\qq
\CP_{_{\CO}}\ =\ \mu^{-1}(\{\CO\})\,/\,\CG\ \cong\ 
\mu^{-1}(\sum\lambda_n\delta_{x_n})\,/\,
\CG_{\underline{\lambda},\underline{x}}
\label{CPG}
\qqq
where $\CG_{\underline{\lambda},\underline{x}}$ is the 
subgroup of $\CG$ fixing $\sum\lambda_n\delta_{x_n}$.
Upon restriction to properly defined stable  
pairs $(A,\Phi)$, $\CP_{_{\CO}}$ gives a smooth space 
with a semi-stable compactification \cite{Mark}.
Its second representation in (\ref{CPG}) allows 
to equip $\CP_{_{\CO}}$ with the symplectic structure 
inherited from $T^*\CA$. The reduced Hamiltonians $h_p$ 
take now values in $H^0(K^{d_p}(d_p\sum x_n))$, i.e. define 
meromorphic $d_p$-differentials with possible poles at $x_n$
of order $\leq d_p$ and they still define an integrable
system on $\CP_{_{\CO}}$. In particular, $h_{p_2}$ takes values
in $H^0(K^2(2\sum x_n))$ which is dual to $H^1(K^{-1}(-2\sum x_n))$,
the tangent space to the moduli space of curves $\Sigma$ with
fixed punctures $x_n$ and first jets at $x_n$ of holomorphic 
local parameters $z_n$, $z_n(x_n)=0$. The corresponding Beltrami 
differentials $\delta\mu$ behave like $\CO(z_n^2)$ around $x_n$ 
and they are taken modulo $\de(\delta\xi)$ where the vector fields 
$\xi$ are also $\CO(z_n^2)$ around $x_n$ (such vector fields
do not change the first jets at $x_n$ of the local parameters 
$z_n$). \m$\delta\mu$ may still be coupled to quadratic 
differentials $\beta\in H^0(K^2(2\sum x_n))$ by (\ref{pair}) 
and Eq.\s\s(\ref{hmu}) defines for $[\delta\mu]\in 
H^1(K^{-1}(-2\sum x_n))$ Hamiltonians on $\CP_{_{\CO}}$ 
that are in involution. 
\vskip 0.5cm

\nsection{Knizhnik-Zamolodchikov-Bernard connection}

\noindent The phase space $\CP\cong T^*\CN$ may be 
(geometrically) quantized by considering the space $H^0({\CL}^k)$ 
of holomorphic sections of the $k^{\rm th}$ power of the determinant 
line bundle ${\CL}$ over $\CN$ (more exactly, over its semi-stable 
version $\CN_{ss}$) as the space of quantum states. Such sections
are given by holomorphic functions $\psi$ on $\CA$ satisfying 
the Ward identity
\qq
\label{CS}
 \psi(A)=\ee^{-k\,S(h,A)}\,\psi({}^{h^{-1}}\hspace{-0.15cm}A)
\qqq
for $h\in\CG$ and with $S(h,A)$ standing for the action of the 
gauged Wess-Zumino-Novikov-Witten (WZNW) model.
The identity (\ref{CS}) expresses the gauge invariance 
on the quantum level. 
The vector spaces $H^0({\CL}^k)$ arise naturally in the context
of the WZNW model and of the Chern-Simons theory \cite{WittJon}.
They are finite-dimensional and their dimension is given 
by the Verlinde formula \cite{Verl}. Put together for different
complex structures of $\Sigma$, they form a holomorphic vector 
bundle $\CW$ over the moduli space of complex curves. 
In the language of functions $\psi$, the $\de$-operator 
of this bundle is given by
\qq
\de_{\overline{\delta\mu}}\psi\,=\,
\left(d_{\overline{\delta\mu}}\,+\,{_k\over^{4\pi i}}
\int_{\Sigma}\tr\,\m (A\m\overline{\delta\mu})\, A\right)\psi
\qqq
where $d_{\overline{\delta\mu}}$ differentiates $\psi$
viewed as a function of the unitary gauge field 
$B=-A^*+A=-A_{\bar z}^*dz+A_{\bar z}d\bar z$ 
(functions of $B$ are naturally identified 
for different complex structures on $\Sigma$). The bundle
$\CW$ may be equipped with a projectively flat connection 
$\nabla^{\rm KZB}$ \cite{WittJon}\cite{Hitch1} which may be traced 
back to the works of Knizhnik-Zamolodchikov \cite{KZ} 
and Bernard \cite{Bern1}\cite{Bern2}. In the present 
description of $\CW$, the KZB connection takes the form 
\cite{Karp}
\qq
&&\nabla^{\rm KZB}_{\delta\mu}\psi\,=\,\left(d_{\delta\mu}\,-\,
	\int_\Sigma\tr\,A^*\m(
	\frac{_\delta}{^{\delta A}}\,\delta\mu)
	\,-\,\frac{_{\pi i}}{^\kappa}
	\int_\Sigma\tr\ {}^\bullet_\bullet\,
	\frac{_\delta}{^{\delta A}}
	(\frac{_\delta}{^{\delta A}}\,\delta\mu)
	\m{}^\bullet_\bullet\right)\psi\, ,\label{KZB}\\
	\cr
	&&\nabla^{\rm KZB}_{\overline{\delta\mu}}\psi\ =\ 
	\de_{\overline{\delta\mu}}\,\psi
	\qqq
where $\kappa=k+g^\vee$ with $g^\vee$ denoting the dual Coxeter 
number of $G$. The symbol ${}^\bullet_\bullet\ {}^\bullet_\bullet$  
indicates that one should remove the singularity at the coinciding 
points of $\frac{\delta}{\delta A(x)}\frac{\delta}{\delta A(y)}\psi$
before setting $x=y$. How this is precisely done depends on some 
choices (e.g. of a projective connection or a metric on each 
$\Sigma$) but the choices lead to connections differing
by addition of a scalar form. 
\vskip 0.4cm

The second order operator on the right hand side of 
Eq.\s\s(\ref{KZB}) has the principal symbol (obtained by 
replacement of $\frac{\delta}{\delta A}$ by
${k\over2\pi i}\Phi$) proportional to the Hitchin Hamiltonian 
$h_{\delta\mu}$. The KZB connection $\nabla^{\rm KZB}_{\delta\mu}$
may be considered a quantization of ${k\over{2\pi i}}h_{\delta\mu}$ 
which, instead of acting in a fixed space $H^0({\CL}^k)$ relates 
two such spaces for the complex structures differing 
by $\delta\mu$\cite{Hitch1}. Note that if we rescale 
$\delta\mu\mapsto\kappa\delta\mu$,
we should obtain from the KZB connection 
in the limit $\kappa\to0$ operators acting in the space 
$H^0({\CL}^{-g^\vee})$ corresponding to a fixed complex structure. 
This space becomes non-trivial if we admit singular sections
of $\CL$ or work with higher cohomologies of $\CL^{-g^\vee}$. 
It also admits a quantization of the non-quadratic Hitchin 
Hamiltonians \cite{BeilDr}\cite{Frenk}. For $k\not=-g^\vee$
we may also obtain from Eq.\s\s(\ref{KZB}) operators
in a single space if we chose a local trivialization 
of the bundle $\CW$ (or of a bundle $\CW'\supset\CW$).
\vskip 0.4cm

The above quantization extends to the case of the phase space
$\CP_{_{\CO}}$ if the coadjoint orbits $\CO_n$ associated
to points $x_n\in\Sigma$ correspond to irreducible holomorphic 
representations of $G$ in vector spaces $V_n$ (i.e. to irreducible 
unitary representations of the compact form of $G$). 
The quantum states are now represented by holomorphic maps 
on $\CA$ with values in $V=\mathop{\otimes}\limits_nV_n$ 
satisfying the Ward identities
\qq
\label{CS1}
 \psi(A)=\ee^{-k\,S(h,A)}\,\mathop{\otimes}\limits_nh(x_n)
 \,\psi({}^{h^{-1}}\hspace{-0.15cm}A)
\qqq
for $h\in\CG$ generalizing Eq.\s\s(\ref{CS}). The spaces 
of solutions are still finite-dimensional and form a holomorphic
vector bundle over the moduli space of punctured curves with first 
jets of local parameters at the punctures. The complex structure 
and the KZB connection are given by the same formulae with
Beltrami differentials $\delta\mu$ restricted to behave 
like $\CO(z_n^2)$ at the punctures. Since $\frac{\delta}
{\delta A(x)}\psi=\CO(z_n^{-1})$ and $\tr\, {}^\bullet_\bullet
\frac{\delta}{{\delta A(x)}}\frac{\delta}{{\delta A(x)}}
{}^\bullet_\bullet\m\psi=\CO(z_n^{-2})$\m\ around $x_n$,
there is no problem of convergence of the integrals over $\Sigma$. 
Again, $\nabla^{\rm KZB}_{\delta\mu}$ may be viewed as the quantization
of the Hitchin Hamiltonian ${k\over{2\pi i}}h_{\delta\mu}$ and 
all the above remarks apply.
\vskip 0.5cm

\nsection{Genus zero}

\noindent Up to diffeomorphisms, there is only one 
Riemann surface of genus zero: the Riemann sphere $\NP^1
=\NC\cup\{\infty\}$. On $\NP^1$, the gauge orbit of the zero gauge 
field is open and dense in $\CA$, i.e. the generic gauge field 
takes the form
\qq
A=h^{-1}\de h
\qqq
where $h\in\CG$ is determined up to left multiplication by
a constant $g\in G$. The equation $\mu(A,\Phi)=
\sum_n\lambda_n\delta_{z_n}$, with $\lambda_n$ belonging
to the (co)adjoint orbit $\CO_n$ associated to the puncture 
$z_n$, becomes
\qq
\de({}^h\Phi)=\sum_n\nu_n\delta_{z_n}
\qqq
where $\nu_n=h(z_n)\lambda_n h(z_n)^{-1}\in\CO_n$. 
This equation has a (unique) solution 
\qq
{}^h\Phi(z)\,=\,\sum_n\frac{\nu_n}{z-z_n}\,\frac{_{dz}}{^{2\pi i}}
\qqq
if and only if the sum of residues is zero, i.e.\s\s if 
$\sum_n\nu_n=0$. We obtain then for $\CP_{_{\CO}}$ defined 
by Eq.\s\s(\ref{CPG}):
\qq
       \CP_{_{\CO}}\,\cong\,
       \Big\{\underline{\nu}\in\mathop{\times}\limits_n\CO_n
       \,\Big|\,\sum_n\nu_n=0\Big\}\bigg/ {G}\,.
\qqq
The (co)adjoint orbits carry a natural symplectic structure
leading to the Poisson bracket $\{\nu^a,\nu^b\}\,
=\,i\m f^{abc}\m\nu^b$ for $\nu^a=\tr\, t^a\nu$ where $t^a$ 
are the generators of $Lie(G)$ s.t. $\tr\, t^at^b=\delta^{ab}$ 
and $[t^a,t^b]=if^{abc}t^c$.
It is easy to check that the complex symplectic structure 
on $\CP_{_{\CO}}$ coincides with the one obtained by 
the symplectic reduction of $\times_n\CO_n$ with respect 
to the diagonal action of ${G}$. 
\vskip 0.4cm

The Hamiltonians $h_p$ are  
\qq
h_{p}(z)=p\Big(\sum_n\frac{\nu_n}
	{z-z_n}\Big)\,(\frac{_{dz}}{^{2\pi i}})^{d_p}.
\qqq
In the special case $p_2={1\over2}\tr$, 
\qq
h_{p_2}\,=\,{_1\over^2}\sum_{n,m}
	\frac{\nu_n^a\nu_m^a}{(z-z_n)(z-z_m)}
	\,(\frac{_{dz}}{^{2\pi i}})^2\,=\,
	\sum_n
	\left(\frac{1}{(z-z_n)^2}\,\delta_n
	+\frac{1}{z-z_n}\,h_n
	\right)(\frac{_{dz}}{^{2\pi i}})^2
\label{h21}
\qqq
where 
\qq
\delta_n={_1\over^2}\m\nu^a_n\nu^a_n\,,\qquad
h_n=\sum_{m\neq n}\frac{\nu^a_n\nu^a_{m}}{z_n-z_{m}}\,.
\qqq
Let $\delta\mu$ be a Beltrami differential regular at infinity
and behaving like $\CO((z-z_n)^2)$ around the insertions.
Necessarily, $\delta\mu=\de(\delta\xi)$ for a regular vector
field $\delta\xi=\delta\xi^z\da_z$ on $\NP^1$. $\delta\xi$
is determined up to infinitesimal M\"{o}bius transformations
$(a+bz+cz^2)\da_z$. We may take $z'=z+\delta\xi^z$ as the new 
complex coordinate on $\NP^1$ with the modified complex structure. 
The modification is then equivalent to the shift $\delta z_n=\delta
\xi^z(z_n)$ of the insertion points and the shift 
$\delta\chi_n=\chi_n\m\da_z(\delta\xi_n^z)(z_n)$ of the first jet 
of the local parameter at the punctures parametrized 
by the $\da_z$-derivative $\chi_n$ of the parameter at $z_n$. 
An easy calculation involving cutting out small balls
around the insertions and integration by parts shows that
\qq
h_{\delta\mu}\ =\ \sum_n\int_\Sigma
	\left(\frac{1}{(z-z_n)^2}\,\delta_n
	+\frac{1}{z-z_n}\,h_n
	\right)(\frac{_{dz}}{^{2\pi i}})^2\,\,\de(\delta\xi)\cr
	=\ {_1\over^{2\pi i}}\sum_n
	\left(\delta_n\m\,\chi_n^{-1}\delta\chi_n
	\,+\, h_n\,\delta z_n\right).
\label{hmu0}
\qqq
\vskip 0.4cm

The quantum states $\psi$ at genus zero may be
labeled by their values $\psi(0)$ at $A=0$
which belong to the subspace $V^G$ of the $G$-invariant 
tensors in $V\equiv\mathop{\otimes}_n V_n$. Indeed, 
$\psi$ is determined by its values on the dense $\CG$-orbit 
of $A=0$. Hence the bundle $\CW$ is a subbundle of the trivial 
bundle with the fiber $V^G$. The KZ(B) connection reduces 
in this case \cite{KZ} to the formula
\qq
&&\nabla^{\rm KZ}_{\delta\mu}\psi(0)\ =\ 
\sum\limits_n\left(\delta\chi_n(\da_{\chi_n}-\chi_n^{-1}\Delta_n)
\,+\,\delta z_n(\da_{z_n}-{_1\over^{\kappa}}\m H_n)
\right)\psi(0)\, ,\label{KZ01}\\
&&\nabla^{\rm KZ}_{\overline{\delta\mu}}\psi(0)\ =\ 
\sum\limits_n\left(\overline{\delta\chi}_n\da_{\bar\chi_n}
\,+\,\overline{\delta z}_n\da_{\bar z_n}
\right)\psi(0)
\label{KZ02}
\qqq
where 
\qq
\Delta_n={_1\over^{2\kappa}}\m t^a_nt^a_n\, ,\quad\quad
H_n\,=\,\sum\limits_{m\not=n}\frac{t_n^at_m^a}{z_n-z_m}\s
\label{H0} 
\qqq
with $t^a_n$ denoting the action of the generator $t^a$ in
the factor $V_n$ of $V$. $\Delta_n$ is a number, the 
{\it conformal weight} assigned in the WZNW theory
to the irreducible representation of $G$ in $V_n$ \cite{KZ}.
Note that, modulo the shift $k\mapsto\kappa=k+g^\vee$,
$\Delta_n$ and ${1\over\kappa}H_n$ may be obtained from 
${k\over(2\pi i)^2}\delta_n$ 
and ${1\over(2\pi i)^2}h_n$, respectively, by the (geometric) 
quantization of the coadjoint orbits which replaces the functions 
$\nu^a_n$ by the operators ${2\pi\over ki}t^a_n$ so that 
the Poisson bracket turns into ${ki\over2\pi}$ times the commutator 
(${2\pi\over k}$ plays the role of the Planck constant).
The flatness of the connection $\nabla^{{\rm KZ}}$,
\m$(\nabla^{{\rm KZ}})^2=0$, follows
from the equation $[H_n,H_m]=0$, equivalent to the classical 
Yang-Baxter equation (CYBE) for ${t^a_1t^a_2\over z}$:
\qq
\Big[\frac{t^a_n\,t^a_{m}}{z_n-z_{m}}\,,\,
	\frac{t^a_{m}\,t^a_{p}}{z_{m}-z_{p}}\Big]+
	{\rm cyclic\ permutations}\,=\,0\, .
	\label{CYBE}
\qqq
\vskip 0.5cm

\nsection{Genus one}

\noindent At genus one, every Riemann surface is isomorphic 
to an elliptic curve $E_\tau\equiv\NC/(\NZ+\tau\NZ)$ 
where $\tau$ is a complex number of positive imaginary part $\imtau$. 
Denote by $\Delta$ ($\Delta_+$) the set of (positive) roots of $Lie(G)$,
by $e_\alpha$ the step generators attached to the roots $\alpha$ 
and let $(\eta^j)$ be an orthonormal basis of the Cartan algebra $Lie(T)$. 
We set $u_\alpha=\tr\,u\alpha$ and $u^j=\tr\,u\eta^j$ for $u\in Lie(T)$.
We shall need some elliptic functions: the Jacobi theta function
\qq
\vartheta_1(z)=-i\sum\limits_{\ell=-\infty}^\infty
(-1)^\ell\,\ee^{\pi i\tau
(\ell+\frac{1}{2})^2+\pi i z\,(2\ell+1)}\, ,
\qqq
the Green function $P_x$ of the twisted $\de$-operator
\qq
P_x(z)=\frac{\vartheta_1'(0)\,\vartheta_1(x+z)}{\vartheta_1(x)\,
\vartheta_1(z)},
\qqq
with the properties $P_x(z+1)=P_x(z)$, $P_x(z+\tau)=\ee^{-2\pi i x}
P_x(z)$ and $P_x(z)={1\over z}+\CO(1)$ around $z=0$,
the function 
\qq
\rho=\vartheta_1'/\vartheta_1
\qqq
s.t. $\rho(z+1)=\rho(z)$, $\rho(z+\tau)=\rho(z)-2\pi i$, 
$\rho(z)={1\over z}+\CO(1)$ around $z=0$
and, finally, the Weyl-Kac denominator
\qq
\Pi(u)=\ee^{2\pi i\tau d/24}
	\prod_{\alpha\in\Delta_+}\,(\ee^{\pi iu_\alpha}-
	\ee^{-\pi iu_\alpha})\,
	\prod\limits_{\ell=1}^\infty\Big[(1-\ee^{2\pi 
	i\ell\tau})^{r}
	\,\prod_{\alpha\in\Delta}(1-\ee^{2\pi 
	i\ell\tau}\ee^{2\pi iu_\alpha})
	\Big]
\qqq
where $d$ denotes the dimension and $r$ the rank of $G$.
\vskip 0.4cm

On $E_\tau$ a generic gauge field is in the orbit of $A_u=
\pi u\,d\bar{z}/\imtau$, for $u\in{Lie(T)}$, i.e.
\qq
A={}^{h^{-1}}\hspace{-0.15cm}A^{01}_u=(h_uh)^{-1}\de(h_uh)
\qqq
where $h_u=\ee^{\pi(u\bar{z}-\bar{u}z)/\imtau}$. Consequently, 
the gauge fields may be parametrized by $u$ and $h$. To avoid 
ambiguities, we have to identify the pairs as follows
\qq
(u,\m h)\,\sim\,(wuw^{-1},\, wh)\,\sim\,(u+q^\vee,\,
h^{-1}_{q^\vee}\,h)\,
\sim
	\,(u+\tau q^\vee,\, h^{-1}_{\tau q^\vee}\,h),
\qqq
for $q^\vee$ in the coroot lattice $Q^\vee$ and $w$ in the 
the normalizer $N$ of ${Lie(T)}$ in $G$. Similarly to 
the genus zero case, we have to solve the equation
\qq
\de\,({}^{h_uh}\Phi)\,=\,\sum_n\nu_n\delta_{z_n}
\qqq
where $\nu_n=(h_uh)(z_n)\,\lambda_n\,(h_uh)(z_n)^{-1}$. 
Decomposing $\nu_n=\sum_\alpha\nu_n^{-\alpha}e_\alpha
+\nu^0_n$ with $\nu^0_n=\nu_n^j\eta^j\in{Lie(T)}$, we can solve 
the above equation if and only if $\sum_n\nu^0_n=0$. 
In that case,
\qq
{}^{h_u h}\Phi(z)\ =\ \left(\varphi_0
	\,+\,\sum_n\Big(\sum_\alpha P_{u_\alpha}(z-z_n)\,
	\nu_n^{-\alpha}e_\alpha
	+\rho(z-z_n)\,\nu^0_n\Big)\right)\frac{_{dz}}{^{2\pi i}}
\qqq
for an arbitrary constant $\varphi_0=\varphi^j_0\m\eta^j$. 
Performing the symplectic reduction, we find
\qq
\CP_{_{\CO}}\,\simeq\,
	\Big\{\m(u,\varphi_0,\underline{\nu})\in T^*{Lie(T)}\times
	(\mathop{\times}\limits_n\CO_n)\,\Big|\,\sum_n\nu^0_n=0\Big\}
	\bigg/ N{\dl o} (Q^\vee+\tau Q^\vee)
\qqq
where the action of $N{\dl o} (Q^\vee+\tau Q^\vee)$ implements
the identifications
\begin{eqnarray*}
(u\m,\s\varphi_0\m,\s\underline{\nu}) & \sim & 
	(wuw^{-1}\m,\s w\varphi_0w^{-1}\m,\s w
	\underline{\nu}w^{-1})\,\sim\,
	\left(u+q^\vee,\s\varphi_0\m,\s
	(h^{-1}_{q^\vee}(z_n)\m\nu_n\m h_{q^\vee}(z_n))\right)\\
	& \sim &
	\left(u+\tau q^\vee\m,\s\varphi_0\m,\s
	(h^{-1}_{\tau q^\vee}(z_n)\,\nu_n\,h_{\tau q^\vee}(z_n))
	\right).
\end{eqnarray*}
The symplectic structure of $\CP_{_{\CO}}$ is that of the reduction
of \m$T^*{Lie(T)}\times(\mathop{\times}\limits_n\CO_n)$ by
the group $N{\dl o} (Q^\vee+\tau Q^\vee)$.
Now it is easy to write down the Hitchin Hamiltonians. Let us 
us do it for $p_2={1\over 2}\tr$. A straightforward computation 
identifying the pole terms leads to
\qq
h_{p_2}\ =\ \Big\{
	-\sum_n\rho'(z-z_n)\,\delta_n
	\,+\,\sum_n\rho(z-z_n)\,h_n\,+\,h_0
	\Big\}\Big(\frac{_{dz}}{^{2\pi i}}\Big)^2
\qqq
where, as before, $\delta_n={1\over 2}\nu^a_n\nu^a_n$ and  
\qq
&&h_0={_1\over^2}\sum_{j=1}^r\varphi_0^j\varphi_0^j
	+{_1\over^2}\sum_{m,n}\Big\{
	\sum_\alpha\partial_xP_{u_\alpha}(z_n-z_m)
	\m\nu^\alpha_n\nu^{-\alpha}_m
	+\frac{_1}{^2}
	\sum_{j=1}^r \frac{\vartheta_1''}{\vartheta_1}(z_n-z_m)
	\m\nu^j_n\nu_m^j\Big\},\hspace{1cm}\\
&&h_n=\sum_{j=1}^r \nu^j_n\varphi_0^j\,+\,\sum_{m\neq n}\Big(
	\sum_\alpha P_{u_\alpha}(z_n-z_m)\,\nu_n^\alpha\nu_m^{-\alpha}
	\,+\,\sum_{j=1}^r\rho(z_n-z_m)\,\nu^j_n\nu^j_m\Big).
\qqq
Note the similarity to the genus 0 case (\ref{h21}).
\vskip 0.4cm

Let $\delta\mu=\delta\mu_{\bar z}^z\m\da_zd\bar z$ 
be a Beltrami differential on $E_\tau$
behaving like $\CO((z-z_n)^2)$ around the insertions.
The modified complex structure corresponds to the
complex coordinate $z'=z+{z-\bar z\over2i\tau_2}
\m\delta\tau\m+\m{\delta\xi}^z$ 
s.t. $\da_{\bar z}z'=\delta\mu_{\bar z}^z$. 
We require that $\delta\xi^z(z+1)=\delta\xi^z(z+\tau)
=\delta\xi^z(z)$. $\delta\tau$ is determined from the condition
that the integral of $\delta\mu_{\bar z}^z$ over
$E_\tau$ is equal to that of ${i\over 2\tau_2}\delta\tau$.
${\delta\xi}^z$ is unique up to an additive constant.
Note that $z'(z+1)=z'(z)+1$ whereas $z'(z+\tau)=z'+\tau'$
where $\tau'=\tau+\delta\tau$. Hence the deformed
curve is isomorphic to $E_{\tau'}$ with the punctures
moved to $z'_n=z_n+\delta z_n$ and the first jets
of local parameters changed to $\chi'_n=\chi_n+\delta\chi_n$
with
\qq
\delta z_n={z_n-\bar z_n\over2i\tau_2}\m\delta\tau\m
+\m{\delta\xi}^z(z_n)\,,\quad\quad\chi_n^{-1}\delta\chi_n
={\delta\tau\over2i\tau_2}\m+\m\da_z\delta\xi^z(z_n)\,.
\qqq
Again by a straightforward calculation substituting
$\delta\mu_{\bar z}^z=\da_{\bar z}z'$, cutting out
small balls around points $z_n$ and integrating by parts,
we obtain
\qq
h_{\delta\mu}\,=\,\int_{E_\tau}h_{p_2}\m\delta\mu\ 
	=\ {_1\over^{2\pi i}}\sum_n
	\left(\delta_n\,\chi_n^{-1}\delta\chi_n
	\,+\, h_n\,\delta z_n\,+\,{_1\over^{2\pi i}}\, 
	h_0\,\delta\tau\right).
\label{hmu1}
\qqq
\vskip 0.4cm

The quantum states $\psi$ at genus one may be characterized
by giving holomorphic functions $\tilde\psi(u)$ on $Lie(T)$ 
with values in $V^T$, the subspace of $T$-invariant tensors
in the product $V$ of the representation spaces,
\qq
\tilde\psi(u)\,=\,\Pi(u)\,\ee^{-\pi k\,\tr\,u^2/(2\imtau)}
\mathop{\otimes}\limits_n \Big(\ee^{-\pi(z_n-\bar{z}_n)
u/\imtau}\Big)_{n}\psi(A_u)\,.
\label{st1}
\qqq
The KZB connection takes the form \cite{EtinK,FaGa,FeWie,FeVa,GaFal}
\qq
&&\nabla^{\rm KZB}_{\delta\mu}\tilde\psi\,=\, 
\sum\limits_n\left(\delta\chi_n(\da_{\chi_n}-\chi_n^{-1}\Delta_n)
\,+\,\delta z_n(\da_{z_n}-{_1\over^{\kappa}}\m H_n)
\,+\,\delta\tau(\da_{\tau}-{_1\over^{2\pi i\kappa}}\m H_0)
\right)\tilde\psi\,,\hspace{1cm}\label{KZB1}\\
&&\nabla^{\rm KZB}_{\overline{\delta\mu}}\tilde\psi\,=\, 
\sum\limits_n\left(\overline{\delta\chi}_n\da_{\bar\chi_n}
\,+\,\overline{\delta z}_n\da_{\bar z_n}\,+\,\overline{\delta\tau}
\m\da_{\bar\tau}\right)\tilde\psi
\label{KZB2}
\qqq
where $\Delta_n$ is as before and the operators
${1\over k}H_0$ and ${1\over k}H_n$ are obtained 
from the Hamiltonians ${k\over(2\pi i)^2}h_0$ and 
${k\over(2\pi i)^2}h_n$ by the replacement
\qq
\varphi_0^j\mapsto {_{2\pi}\over^{ki}}\partial_{u^j},\qquad
	\mu^\alpha_n\mapsto {_{2\pi}\over^{ki}}e_{\alpha\m n},
	\qquad \mu^j_n\mapsto {_{2\pi}\over^{ki}}\eta^j_n\,,
\qqq
i.e. by the geometric quantization.
The resulting Hamiltonians $H_0$ and $H_n$ act on general 
meromorphic functions on $Lie(T)$ with values in $V^T$.
The flatness of the KZB connection is ensured by their 
commutation: $[H_n,H_m]=0$, for $n,m=0,1,\dots$, following from
the so called dynamical CYBE~\cite{FeWie}. 
\vskip 0.5cm

\nsection{Genus two}
\subsection{Curve and its Jacobian}

\noindent Let $\Sigma$ be a curve of genus 2. Choosing a marking,
i.e.\s\s a symplectic homology basis $(A^a,B^a),\ a=1,2,$ on $\Sigma$,
we may fix the corresponding basis $(\omega^a)$ of the
holomorphic 1,0-forms (abelian differentials)
s.t. $\int_{A^a}\omega^b=\delta^{ab}$. 
The $B^a$-periods of the abelian differentials give rise 
to the symmetric period matrix $\tau$, $\tau^{ab}=
\int_{B^a}\omega^b$, with a positive imaginary part. 
The map
\qq
\Sigma\ni x\,\mapsto\,z(x)={\omega^2(x)\over\omega^1(x)}
\label{doco}
\qqq
realizes $\Sigma$ as a double covering of $\NP^1$
ramified over six Weierstrass points $x_n$ or as
a hyperelliptic curve given by the equation
\qq
y^2=\prod_{n=1}^6(z-z_n)\,.
\label{curve}
\qqq
The coordinates $z_n=z(x_n)$ are assumed to be finite. 
This may be always achieved by an appropriate choice 
of the marking. Curve $\Sigma$ may be viewed as composed 
of the points $(y,z)$ and of two points at infinity. 
The covering of $\NP^1$ is $(y,z)\mapsto z$. In this 
representation, the holomorphic 1,0-forms and the 
holomorphic quadratic differentials on $\Sigma$ 
have the form
\qq
\omega=(a+bz)dz/y\,,\qquad \beta=(a+bz+cz^2)(dz)^2/y^2\,,
\label{qdiff}
\qqq
respectively. In particular,
\qq
\omega^1\ \propto\ dz/y\,,\qquad\omega^2
\ \propto\ z\m dz/y\,
\qqq
with the same proportionality constant.
\vskip 0.4cm

The Jacobian $J^1$ of the degree 1 holomorphic line bundles 
may be represented as $\NC^2/(\NZ^2+\tau\NZ^2)$. 
In particular, the spin structures $L$, $L^2=K$, correspond
to points $e+\tau e'$ with $e\in{1\over 2}\NZ^2/\NZ^2$.
There are 6 odd spin structures $e_n+\tau e'_n$ 
labeled by the Weierstrass points $z_n$, the zeros
of the holomorphic sections they admit, and 10 
even spin structures without holomorphic sections.
\vskip 0.2cm

\subsection{Theta functions}

\noindent The degree 1 bundles with holomorphic sections form 
the theta-divisor in $J^1$. The holomorphic sections of the 
$k^{\rm th}$-power of the corresponding theta bundle
over $J^1$ may be represented by the holomorphic theta
functions of degree $k$ on $\NC^2$ defined by the relations
\qq
\ee^{\m\pi i k\s n\cdot\tau n\m+\m
2\pi i k\s n\cdot u}\,\theta(u+m+\tau n)\,=\,\theta(u)
\label{theta}
\qqq
for $m,n\in\NZ^2$. They form the space $\Theta_{k}$ of dimension $k^2$. 
For $k=1$ there is a single (up to normalization) 
theta function 
\qq
\vartheta(u)\,=\,\sum\limits_{n\in\NZ^2}\ee^{\m\pi i\s n\cdot\tau n
+2\pi i\s n\cdot u}\,,
\qqq
the Riemann theta function.
For $k=2$ there are four independent $\theta$-functions.
One can take them as
\qq
\theta_{e}(u)\,=\,\sum\limits_{n\in\NZ^2}\ee^{\m2\pi i\s (n+e)\cdot
\tau(n+e)\m+\m4\pi i\s (n+e)\cdot u}
\label{base}
\qqq
for $e\in{1\over 2}\NZ^2/\NZ^2$. 
\qq
\vartheta(u+v)\m\vartheta(u-v)\s=\s\sum\limits_e\theta_e(u)\m
\theta_e(v)
\label{thth}
\qqq
is a second order theta function in both $u$ and $v$.
The map \s$v\,\mapsto\,\vartheta(\s\cdot\s+v)\m\vartheta
(\s\cdot\s-v)\s$ determines an embedding of the Kummer 
surface $J^1/\NZ_2\cong \NC^2/(\NZ^2+\tau\NZ^2)/\NZ_2$ 
onto a quartic surface $\CK$ in the 3-dimensional projective 
space $\NP\Theta_2\s$ ($\NZ_2$ maps the degree 1 line bundles 
$L$ into $L^{-1}K$ or $v\in\NC^2$ into $-v$).
\vskip 0.4cm

The double theta function (\ref{thth}) determines a non-degenerate
symmetric quadratic form on the space $\Theta_2^*$ dual to $\Theta_2$.
It permits to identify $\Theta_2^*$ with $\Theta_2$ by
sending $\phi\in\Theta_2^*$ to $\iota(\phi)\in\Theta_2$ defined by
\qq
\iota(\phi)(u)\,=\,\langle\m\vartheta(u+\s\cdot\m\s)\m
\vartheta(u-\s\cdot\m\s)\m,\s\phi\m\rangle\s.
\label{iota}
\qqq
The identification exchanges the basis $(\theta_e)$ of $\Theta_2$ 
with the dual basis $(\theta_e^*)$ and the Kummer quartic $\CK$
with its dual version $\CK^*\subset\NP\Theta_2^*$. $\CK^*$ 
is composed of linear forms proportional to the evaluation
forms $\phi_u$ defined by
\qq
\langle\theta,\m\phi_u\rangle\,=\,\theta(u)\,.
\qqq
\vskip 0.4cm

The group $({1\over^2}\NZ/\NZ)^4$ of spin structures acts 
on $\Theta_k$ for even $k$ by endomorphisms $U_{e,e'}$ 
defined by 
\qq
(U_{e,e'}\m\theta)(u)\,=\,\ee^{\m\pi i k\s e'\cdot\tau e'\m+\m
2\pi i k\s e'\cdot u}\,\theta(u+e+\tau e')\,.
\qqq
For $k$ not divisible by $4$ this action is only projective:
$U_{e,e'}U_{f,f'}=(-1)^{4\s e\cdot f'}\, U_{e+f,\m e'+f'}$ and
it lifts to a Heisenberg group. For $k=2$,
\qq
U_{e,e'}\m\theta_{e''}=(-1)^{4\s e\cdot e''}\,\theta_{e'+e''}\,.
\label{act1}
\qqq
The action of \s$U_{e,e'}$ preserves the Kummer quartic $\CK\subset
\NP\Theta_2$ and the action of the transposed endomorphisms 
$U_{e,e'}^{\ t}$ preserves $\CK^*$.
\vskip 0.2cm

\subsection{Moduli space of $SL_2$-bundles}

\noindent In the fundamental paper~\cite{NarRa}, Narasimhan 
and Ramanan proved that the  moduli space $\CN_{s}$ of the stable 
$SL_2$ holomorphic bundles is canonically isomorphic to $\NP\Theta_2
\setminus\CK$. The isomorphism associates to an $SL_2$-bundle $E$
the second order theta function $\theta$ vanishing at the points 
$u\in\NC^2$ 
corresponding to the duals of the line-subbundles of the rank 2 
bundle associated to $E$. In other words, if $A$ is the gauge
field whose $\CG$-orbit corresponds to $E$ and if $L_u$ denotes 
the degree 1 line bundle corresponding 
to $u\in\NC^2/(\NZ^2+\tau\NZ^2)$ then
the theta function $\theta$ associated to $E$ vanishes at $u$
if and only if there exists a pair $s=(s_1,s_2)$ composed
of sections of $L_u$ s.t. $(\de+A)s=0$. The semistable 
compactification of the moduli space $\CN_{s}$ is
\qq
\label{thetaiso}
\CN_{ss}\ \cong\ \NP\Theta_2
\qqq
and, exceptionally, it is smooth. The points of the Kummer quartic 
$\CK$ represent (classes of) the semistable but not stable bundles. 
Hence for $G=SL_2$ the phase space of the Hitchin system on the genus 
2 curve with no insertions is 
\qq
T^*\CN_{ss}\ \cong\ T^*\NP\Theta_2\ \cong
\ \{\s(\theta,\phi)\in\Theta_2\times\Theta_2^{\m*}
\s\s\s|\s\s\s \theta\not=0,\ \langle\theta,\phi\rangle=0\s\}
\Big/\NC^\times
\qqq
with the action of $t\in\NC^\times$ given by $(\theta,\phi)
\mapsto(t\m\theta,\m t^{-1}\phi)$. As a symplectic space,
it is the symplectic reduction of $T^*(\Theta_2\setminus\{0\})$
by the action of $\NC^\times$. Using
the bases $(\theta_e)$ and $(\theta_e^*)$ to decompose
\qq
\theta=\sum q_e\theta_e\,,\qquad\phi=\sum p_e\theta_e^*\,,
\label{dec}
\qqq
we may represent $T^*\CN_{ss}$ as the space of pairs
$(q,p)\in\NC^4\times\NC^4$, $q\not=0$, $q\cdot p=0$, 
with the identification $(q,p)\sim(t\m q,\m t^{-1}p)$ 
and the symplectic form induced from $dp\cdot dq$.
\vskip 0.2cm

\subsection{Hitchin map for $G=SL_2$}

The Hitchin map $h_{p_2}\m:\m T^*\NP\Theta_2\rightarrow H^0(K^2)$
appears to take a particularly simple form resembling
the genus 0 formula (\ref{h21}):
\qq
h_{p_2}\ =\ {_1\over^{2}}\sum\limits_{n,m=1\atop n\not=m}^6
{r_{nm}\over(z-z_n)(z-z_m)}\s({_{dz}\over^{2\pi i}})^2
\ =\ \sum\limits_{n=1}^6{h_n\over z-z_n}\s
({_{dz}\over^{2\pi i}})^2\hspace{0.4cm}
\label{h22}
\qqq
where
\qq
r_{nm}(\theta,\m\phi)\ =\ {_1\over^{16}}\s
\langle\m U_{e_n,e'_n}\theta\m,\s
U_{e_m,e'_m}^{\ t}\phi\m\rangle\s
\langle\m U_{e_m,e'_m}\theta\m,\s 
U_{e_n,e'_n}^{\ t}\phi\m\rangle\s
\label{rnm}
\qqq
(the last two factors on the right hand side coincide modulo sign)
and
\qq
h_n\ =\ \sum\limits_{m\not=n}^6{r_{nm}\over{z_n-z_m}}\,.
\label{h2n}
\qqq
With the help of Eq.\s\s(\ref{act1}), $r_{nm}$'s may be
rewritten in the language of $q$'s and $p$'s as explicit 
homogeneous polynomials of order 2 in $q_e$ and in $p_e$,
see below. The identity
\qq
\sum\limits_{m\not=n} r_{nm}\,=\,0
\label{srnm}
\qqq
holding for each $n$ guarantees that 
\s$\prod\limits_{n=1}^6(z-z_n)\sum\limits_{n,m=1\atop n\not=m}^6
{r_{nm}\over(z-z_n)(z-z_m)}\s$ is a quadratic polynomial in $z$.
It follows that the right hand side of Eq.\s\s(\ref{h22}) 
determines a quadratic differential on $\Sigma$ of the general
form given by Eq.\s\s(\ref{qdiff}). 
\vskip 0.4cm

The equality (\ref{h22}) is not immediate. It was established 
in four steps. We shall only enumerate them here. The two first 
crucial steps were performed in \cite{VGP}. It was shown 
there that for any $\theta\not=0$ and $u\in\NC^2$ s.t.  
$\theta(u)=0$,
\qq
h_{p_2}(\theta,\phi_u)\ =\ -{_1\over^{16\pi^2}}\s(\da_{u^a}\theta(u)
\s\omega^a)^2\,.
\label{onKum}
\qqq
The above equation describes the quadratic polynomial 
$h_{p_2}(\theta,\s\cdot\s)$
(with values in $H^0(K^2))$ on the quartic $\CK^*_\theta
=\CK^*\cap\NP\theta^\perp$ in the projectivized subspace 
of $\Theta_2^*$ perpendicular to $\theta$. In principal, 
it determines $h_{p_2}$ completely. It was 
observed then that the above formula implies that for each 
Weierstrass point $z_n$ the conic 
\qq
\CC_n\ =\ \{\m\NC^\times\phi\s\in\s\NP\theta^\perp\ 
\vert\ h_{p_2}(\theta,\phi)\vert_{z_n}=0\m\}
\label{conic}
\qqq
is in fact the union of two bitangents to $\CK_\theta^*$. 
The explicit equations for the bitangents to the Kummer quartics 
known since about a century permitted then to establish 
Eq.\s\s(\ref{h22}) up to multiplication by a $\theta$-dependent
factor \cite{VGP}. The other steps in the proof of the
formula (\ref{h22}) were taken in \cite{GaTr} were it was 
established that the Hitchin map $h_{p_2}$ possesses
the important self-duality property:
\qq
h_{p_2}(\iota(\phi),\m\iota^{-1}(\theta)\m)\ =\ h_{p_2}
(\theta,\m\phi)
\label{sdu}
\qqq
or that $h_{p_2}(q,p)=h_{p_2}(p,q)$ in the language of (\ref{dec}).
This property, far from obvious in the original formulation
of the Hitchin system, restricted the ambiguity on the right
hand side of Eq.\s\s(\ref{h22}) to a (possibly curve-dependent) 
constant factor. The latter was fixed in \cite{GaTr} by a
tedious calculation of $h_{p_2}$ at special points of
$\CK\times\CK^*$.
\vskip 0.2cm

\subsection{Deformations of complex structure}

\noindent For the hyperelliptic curve, 
the variations of the complex structure 
described by the Beltrami differentials $\delta\mu$
on $\Sigma$ change the images $z_n$ 
of the ramification points of the covering of $\NP^1$. 
Let us find these changes. Let
\qq
{\omega'}^a\,=\,\omega^a\s+\s\delta\omega^a+\bar\delta\omega^a
\qqq
denote a deformed basis $({\omega'}^a)$ of the abelian 
differentials with $\delta\omega^a$ of 1,0-type and $\bar\delta
\omega^a$ of 0,1-type in the original complex structure. 
$\delta\omega^a$ and $\bar\delta\omega^a$ have to satisfy
the relations
\qq
\bar\delta\omega^a\,=\,\omega^a\m\delta\mu\,,\qquad
\de(\delta\omega^a)\,=\,-\,\da(\omega^a\m\delta\mu)
\label{sp0}
\qqq
stating, respectively, that ${\omega'}^a$ is of the 1,0-type
in the deformed structure and that it is a closed form.
The equation for $\delta\omega^a$ always has solutions.
They are defined modulo abelian differentials and the ambiguity
may be fixed by demanding that $\int_{A^a}{\omega'}^b=\delta^{ab}$.
The deformed covering map onto $\NP^1$ is then
\qq
z'(x)\,=\,{{\omega'}^2(x)\over{\omega'}^1(x)}\,=\,
z(x)\,+\,{\delta\omega^2\over\omega^1}(x)\,-\,
z(x)\,{\delta\omega^1\over\omega^1}(x)\,.
\label{zpx}
\qqq
The ramification points $x'_n$ of the map $z'$ are determined by
solving the equation
\qq
\da' z'(x'_n)\,=\,0\,
\qqq
which, upon rewriting $x'_n=x_n+\delta x_n$ becomes
\qq
(\da)^2z(x_n)\s\delta x_n\,+\,\da\m({\delta\omega^2\over\omega^1})(x_n)
-\da\m( z\m{\delta\omega^1\over\omega^1})(x_n)\ =\ 0\,.
\qqq
Since $\de\m z(x_n)=0$, the quadratic differential $\m(\da)^2z(x_n)$ 
is well defined. Besides, since the ramification points are
isolated, it does not vanish so that one may solve the above 
equations for $\delta x_n$. The ramification points $x'_n$ are 
mapped to
\qq
z'_n\,=\, z'(x_n+\delta x_n)\,=\, z_n\,+\,{\delta\omega^2\over\omega^1}
(x_n)\,-\, z_n\m{\delta\omega^1\over\omega^1}(x_n)\s.
\qqq
We infer that
\qq
\delta z_n\equiv z'_n-z_n\,=\,{\delta\omega^2\over\omega^1}
(x_n)\,-\, z_n\m{\delta\omega^1\over\omega^1}(x_n)
\label{dzpn}
\qqq
are the variations of $z_n$ corresponding to the Beltrami
differential $\delta\mu$.
\vskip 0.4cm

We may now find the values of the Hitchin Hamiltonians 
$\, h_{\delta\mu}\,=\,\int_\Sigma h_{p_2}\s\delta\mu\,$
related to the Beltrami differentials.
Note that, by virtue of Eqs.\s\s(\ref{zpx}) and (\ref{sp0}),
\qq
\de(\delta z)\,\equiv\,
\de\m(z'-z)\,=\,\,\de\m(\delta\omega^2)\m/\m\omega^1\,-\,
z\,\de\m(\delta\omega^1)\m/\m\omega^1\s=\s
-\m\da\m(\omega^2\m\delta\mu)\m/\m\omega^1\,\,\cr\cr
+\,z\,\da\m(\omega^1\m\delta\mu)\m/\m\omega^1
=\,-\m\da\m(z\s\omega^1\m\delta\mu)\m/\m\omega^1
\,+\,z\s\da\m(\omega^1\m\delta\mu)\m/\m\omega^1
\ =\ (dz)\,\delta\mu\,.
\label{dzpn1}
\qqq
A straightforward integration by parts over the region
in $\Sigma$ without small balls around the Weierstrass
points $x_n$ and around 2 points at infinity gives now:
\qq
h_{\delta\mu}\ =\ \int_\Sigma\sum\limits_{n=1}^6
{h_n\over z-z_n}\s ({_{dz}\over^{2\pi i}})^2\m\delta\mu\ =\ 
({_1\over^{2\pi i}})^2\int_\Sigma\sum\limits_{n=1}^6
{h_n\over z-z_n}\s dz\,\de(\delta z)\cr
\ =\ {_1\over^{\pi i}}\sum\limits_{n=1}^6 h_n\,\delta z_n\,.
\hspace{0.5cm}
\label{h2dm}
\qqq
The comparison with Eq.\s\s(\ref{hmu0}) shows an additional 
factor 2 which comes from the double covering.
\vskip 0.2cm

\subsection{Relation to Neumann systems}

\noindent The Poisson-commutation of the Hamiltonians 
$h_n$ (of which any 3 give independent action variables
of our integrable system) is equivalent to the CYB-like 
equations
\qq
\{{r_{nm}\over z_n-z_m}\m,\,{r_{mp}\over z_m-z_p}\}\,
+\,{\rm cycl.}\,=\,0\,,\qquad\{{r_{nm}\over z_n-z_m}\m,\,
{r_{pq}\over z_p-z_q}\}\,=\,0
\label{CYBL}
\qqq
for $\{n,m\}\cap\{p,q\}=\emptyset$. The above relations may
may be directly checked, as noticed in \cite{VGP}. The more 
recent observations of the paper \cite{VGDJ} 
on the Knizhnik-Zamolodchikov-Bernard connection 
in the same setup (see below) permit to identify 
the integrable system with the Hamiltonians $h_n$ 
of Eq.\s\s(\ref{h2n}): it is a modified version of the classical 
genus 2 Neumann systems \cite{Neum,Mumf,AvTa} whose original 
version is also rooted in the modular geometry of hyperelliptic 
curves. This goes as follows. 
\vskip 0.4cm

The phase space $T^*\NP^3$ (without
the zero section) may be identified with the coadjoint orbit 
$\CO_1$ of the complex group $SL_4$ composed of the traceless 
rank 1 matrices $\vert q\rangle \langle p\vert$. The action 
of $SL_4$ in $\wedge^2\NC^4$ preserves
the quadratic form induced by the exterior product 
on $\wedge^2\NC^4$ and the identification 
$\wedge^4\NC^4\cong\NC^1$. It leads to the double covering
$SL_4\rightarrow SO_6$ if we choose in $\wedge^2\NC^4$ 
the Pl\"{u}cker basis turning the quadratic form into
the sum of squares. The inverse relation is the complexified 
version of the twistor calculus. Upon the identification of 
$sl_4$ with $so_6$, the coadjoint orbit $\CO_1$ becomes 
the one composed of (complex) rank 2 antisymmetric matrices
$J=(J_{nm})$ of square zero: $J^2=0$. Such matrices
are of the form $J_{nm}=Q_nP_m-P_nQ_m$ with vectors $P,Q\in\NC^6$ 
spanning an isotropic subspace, i.e. with $Q^2=Q\cdot P=P^2=0$.
Given $(q,p)$ with $q\cdot p=0$, in order to find 
the corresponding pair $(Q,P)$, it is enough to complete vector 
$q$ to a basis $(q,f_1,f_2,f_3)$ of $\NC^4$ s.t. 
$f_1\cdot p=f_2\cdot p=0$ and $f_3\cdot p=1$ and to set
\qq
Q=q\wedge f_1\,,\quad R=f_2\wedge f_3\,,\quad P=q\wedge 
f_2\m/\m Q\cdot R
\label{PQ} 
\qqq
in the language of $\wedge^2\NC^4\cong\NC^6$. 
In the Pl\"{u}cker coordinates, we may take
\qq
&&\hbox to 4cm{$Q_1\s=\s-(q_1p_3+q_4p_2)\,,$\hfill}\quad
\hbox to 4cm{$Q_2\s=\s i(q_1p_3-q_4p_2)\,,$\hfill}\quad
\hbox to 4cm{$Q_3\s=\s-i(q_1p_2+q_4p_3)\,,$\hfill}\cr\cr
&&\hbox to 4cm{$Q_4\s=\s q_1p_2-q_4p_3\,,$\hfill}\quad
\hbox to 4cm{$Q_5\s=\s q_2p_2+q_3p_3\,,$\hfill}\quad
\hbox to 4cm{$Q_6\s=\s\, i(q_2p_2+q_3p_3)\,,$\hfill}\cr\cr
&&\hbox to 4cm{$P_1\s=\s-{1\over2}\,{{q_2p_4+q_3p_1}
\over{q_2p_2+q_3p_3}}\,,$\hfill}\quad
\hbox to 4cm{$P_2\s=\s{i\over2}\,{{q_2p_4-q_3p_1}
\over{q_2p_2+q_3p_3}}\,,$\hfill}\quad
\hbox to 4cm{$P_3\s=\s{i\over2}\,{{q_3p_4+q_2p_1}
\over{q_2p_2+q_3p_3}}\,,$\hfill}\cr\cr
&&\hbox to 4cm{$P_4\s=\s-{1\over2}{{q_3p_4-q_2p_1}
\over{q_2p_2+q_3p_3}}\,,$\hfill}\quad
\hbox to 4cm{$P_5\s=\s{1\over2}\,,$\hfill}\quad
\hbox to 4cm{$P_6\s=\s-{i\over2}$\hfill}
\nonumber
\qqq
where $q_1\equiv q_{(0,0)}$, \s$q_2\equiv q_{({1\over 2},0)}$,
\s$q_3\equiv q_{(0,{1\over 2})}$, \s$q_4\equiv q_{({1\over 2},
{1\over 2})}$ and similarly for $p$'s.
In terms of $(Q,P)$, the symplectic form is $dP\cdot dQ$.
The functions $J_{nm}$ on $\CO_1$ have the Poisson bracket
\qq
&&\hbox to 6cm{$\{J_{nm},\m J_{mp}\}\ =\ -J_{np}$\hfill}
{\rm for}\ \ n,m,p\ \ \ \ \ {\rm different},\label{soa1}\\
&&\hbox to 6cm{$\{J_{nm},\m J_{pq}\}\ =\ 0$\hfill}{\rm for}
\ \ n,m,p,q\ \ {\rm different}\label{soa2}
\qqq
A straightforward check shows now that
\qq
&&J_{12}\ =\ \ \ \m{_i\over^2}(q_1p_1+q_2p_2-q_3p_3-q_4p_4)\s,\quad
J_{13}\ =\ -{_i\over^2}(q_1p_4-q_2p_3-q_3p_2-q_4p_1)\s,\hspace{1.06cm}\cr
&&J_{14}\ =\ \ \ \m{_1\over^2}(q_1p_4+q_2p_3-q_3p_2-q_4p_1)\s,\quad
J_{15}\ =\ -{_1\over^2}(q_1p_3-q_2p_4-q_3p_1+q_4p_2)\s,\hspace{1.06cm}\cr
&&J_{16}\ =\ \ \ \m{_i\over^2}(q_1p_3+q_2p_4+q_3p_1+q_4p_2)\s,\quad
J_{23}\ =\ -{_1\over^2}(q_1p_4-q_2p_3+q_3p_2-q_4p_1)\s,\hspace{1.06cm}\cr
&&J_{24}\ =\ -{_i\over^2}(q_1p_4+q_2p_3+q_3p_2+q_4p_1)\s,\quad
J_{25}\ =\ \ \ \m{_i\over^2}(q_1p_3-q_2p_4+q_3p_1-q_4p_2)\s,\hspace{1.06cm}
\label{Js}\\
&&J_{26}\ =\ \ \ \m{_1\over^2}(q_1p_3+q_2p_4-q_3p_1-q_4p_2)\s,\quad
J_{34}\ =\ -{_i\over^2}(q_1p_1-q_2p_2+q_3p_3-q_4p_4)\s,\hspace{1.06cm}\cr
&&J_{35}\ =\ -{_i\over^2}(q_1p_2+q_2p_1+q_3p_4+q_4p_3)\s,\quad
J_{36}\ =\ -{_1\over^2}(q_1p_2-q_2p_1-q_3p_4+q_4p_3)\s,\hspace{1.06cm}\cr
&&J_{45}\ =\ \ \ \m{_1\over^2}(q_1p_2-q_2p_1+q_3p_4-q_4p_3)\s,\quad
J_{46}\ =\ -{_i\over^2}(q_1p_2+q_2p_1-q_3p_4-q_4p_3)\s,\hspace{1.06cm}\cr
&&J_{56}\ =\ \ \ \m{_i\over^2}(q_1p_1-q_2p_2-q_3p_3+q_4p_4)\nonumber
\qqq
and that
\qq
r_{nm}\ =\ -{_1\over^4}\,(J_{nm})^2\s.
\label{jsq}
\qqq
The equations (\ref{CYBL}) assuring that the Hamiltonians
\s$h_n=-{1\over 4}\sum\limits_{m\not= n}{J_{nm}^{\s2}\over z_n-z_m}$ 
Poisson-commute follow directly from the 
$so_6$ algebra (\ref{soa1}), (\ref{soa2}). The original
Neumann systems are very similar but involve the
coadjoint orbits of $SO_{N}$ composed from rank 2 antisymmetric
matrices of square $\not=0$ \cite{AvTa}. Such orbits, contrary
to the one we consider, have nontrivial standard real forms. 
\vskip 0.2cm

\subsection{Lax matrix approach}

\noindent Although the change of the orbit modifies dimensional 
counts and many details, the methods used in the analysis 
of the Neumann systems, in particular the Lax method developed
in \cite{AvTa}, generalize with minor variations to
our system and permit to find explicitly the angle variables
of the genus 2 Hitchin system. The Lax matrix may be taken as
$L(\zeta)=(L_{nm} (\zeta))$ with
\qq
L_{nm}(\zeta)\ =\ \zeta\, J_{nm}\,+\, z_n\m\delta_{nm}\,.
\label{lax}
\qqq
As in \cite{AvTa}, the Poisson brackets (\ref{soa1}), (\ref{soa2}) 
may be rewritten in the matrix form as
\qq
\{L(\zeta)\otimes 1\m,\s 1\otimes L(\zeta')\}
\ =\ [L(\zeta)\otimes 1\m,\, r^-(\zeta,\zeta')]
\ -\ [1\otimes L(\zeta')\m,\, r^+(\zeta,\zeta')]
\label{r1r2}
\qqq                                
where the $r$-matrices
\qq
r^{\pm}(\zeta,\zeta')\,=\,{_{\zeta\m\zeta'}\over^{\zeta+\zeta'}}\, 
C\,\pm\,{_{\zeta\m\zeta'}\over^{\zeta-\zeta'}}\, T 
\label{rpm}
\qqq
with $C_{mn,qp}=\delta_{mq}\delta_{np}$ and 
$T_{mn,qp}=\delta_{mp}\delta_{nq}$ \m satisfy the CYBE.
The above form of the Poisson bracket implies immediately that
\qq
\{\m\tr\, L(\zeta)^\ell\m,\m\,\tr\, L(\zeta')^{\ell'}\m\}\ =\ 0
\qqq
for all $\zeta,\m\zeta'$. Since
\qq
{_{d^2}\over^{d^2\zeta}}\,\tr\,L(0)^\ell\ =\ 
2\m\ell\sum\limits_{n,m=1\atop n\not=m}^6z_n^{\s\ell-1}
{J_{nm}^{\s2}\over z_n-z_m}\,,
\qqq
the Hamiltonians 
$h_n=-{1\over 4}\sum\limits_{m\not= n}{J_{nm}^{\s2}\over z_n-z_m}$
may be expressed as combinations of the quantities 
$\tr\, L(\zeta)^\ell$. It is not difficult to see that the converse
is also true.
\vskip 0.4cm

More generally, Eq.\s\s(\ref{rpm})
implies that
\qq
\{\m\tr\, L(\zeta)^\ell\m,\, L(\zeta')\m\}
\ =\ [\m M_\ell(\zeta,\zeta')\m,\, L(\zeta')\m]
\label{M1}
\qqq
with
\qq
M_\ell(\zeta,\zeta')\ =\ \ell\,{_{\zeta\m\zeta'}
\over^{\zeta-\zeta'}}\, L(\zeta)^{\ell-1}\,+\,
\ell\,{_{\zeta\m\zeta'}\over^{\zeta+\zeta'}}\, 
L(-\zeta)^{\ell-1}\,.
\label{M2}
\qqq
It follows that the commuting time evolutions of the Lax 
matrix $L(\zeta)$ generated by the Hamiltonians $h_n$,
\qq
\delta_nL(\zeta)\,
=\,\{\m h_n\m,\, L(\zeta)\m\}\,\delta t_n\,,
\label{dtn}
\qqq
are isospectral. In other words, the spectral curve $\CS$ given 
by the characteristic equation
\qq
\det\m(L(\zeta)-z)\ =\ 0
\label{che}
\qqq
is left invariant by the dynamics generated by any of the
Hamiltonians $h_n$. An easy calculation using the fact that
the matrix $J$ has rank 2 gives
\qq
\det\m(L(\zeta)-z)\ =\ \prod\limits_{n=1}^6(z-z_n)\Big(
1\,+\,{_1\over^2}\m\zeta^2\sum\limits_{n,m=1\atop n\not=m}^6
{J_{nm}^{\s 2}\over(z-z_n)(z-z_m)}\Big)\,.
\label{che1}
\qqq
Upon the substitution $\sigma\m=\m {1\over i\m\zeta}\m
\prod_n(z-z_n)$,
the characteristic equation (\ref{che}) becomes
\qq
\sigma^2\ =\ {_1\over^2}\prod\limits_{n=1}^6(z-z_n)^2
\sum\limits_{n,m=1\atop n\not=m}^6{J_{nm}^{\s 2}
\over(z-z_n)(z-z_m)}\ \equiv\ P(z)\,.
\label{che2}
\qqq
Since $P(z)$ is a polynomial in $z$ of order
8 (see the remark after Eq.\s\s(\ref{srnm})),
this is the equation of a hyperelliptic curve $\CS$
of genus 3 composed of pairs $(\sigma,z)$ and 2 points
$p^\pm_\infty$ corresponding to $z=\infty$. We shall
consider only such points of the phase space that
$\CS$ is smooth. The 1,0-forms 
\qq
\Omega^b\,=\,z^b\, dz\m/\m\sigma
\label{Omb}
\qqq
with $b=0,1,2$ form a basis of the abelian differentials 
on $\CS$.
\vskip 0.4cm

We shall search for the eigenvectors $X=(X_n)$ of the 
Lax matrix.
This will allow to adapt the arguments described in great detail
in Sect.\s\s4 of \cite{Mumf} to the present case.
The eigenvector equations
\qq
\zeta\m J_{nm}\m X_m\ =\ \zeta\m Q_n\,(P\cdot X)
-\zeta\m P_n\,(Q\cdot X)\ =\ (z-z_n)\m X_n
\label{evec}
\qqq
imply that
\qq
(z-z_n)\m X_n\ =\ a\m Q_n\,+\, b\m P_n\s.
\qqq
Upon multiplication by $\sigma\m=\m{1\over^i}
\m\zeta^{-1}\prod_n(z-z_n)$,
Eq.\s\s(\ref{evec}) becomes a system of 2 linear equations
for $a$ and $b$:
\qq
&&(\sigma+V)\s a\m\ +\ \m i\m W\s b\ =\ 0\,,\cr
&&-\m i\m U\s a\,\s+\,\s(\sigma-V)\s b\ =\ 0
\label{s22}
\qqq
where
\qq
&&U(z)\ =\ \ \prod\limits_{n=1}^6(z-z_n)\sum\limits_{n=1}^6
{Q_n^{\s2}\over z-z_n}\s,\cr
&&V(z)\ =\ i\m\prod\limits_{n=1}^6(z-z_n)\sum\limits_{n=1}^6
{Q_nP_n\over z-z_n}\s,\cr
&&W(z)\ =\ \ \prod\limits_{n=1}^6(z-z_n)\sum\limits_{n=1}^6
{P_n^{\s2}\over z-z_n}
\label{UVW}
\qqq
are $4^{\rm th}$-order polynomials in $z$. The non-trivial
solution exists if 
\qq
\sigma^2\ =\ U(z)W(z)+V(z)^2\ =\ P(z)
\qqq
where the last equality follows by a straightforward check.
The system (\ref{s22}) of linear equations defines a holomorphic
line subbundle $L$ of the rank 2 bundle $W=\NC^2\otimes
\CO(4p^+_\infty+4p^-_\infty)$ over the hyperelliptic 
curve $\CS$ (the coefficients behave as $z^4$
at infinity). As solutions of (\ref{s22}) we may take 
for example
\qq
a\,=\,\sigma-V(z)\,,\quad b\,=\, i\, U(z)\,\ \qquad{\rm or}
\ \qquad a\,=\,-\m i\, W(z)\,,\quad b\,=\,\sigma+V(z)\s.
\label{sol}
\qqq
Since $a$ and $b$ are proportional to $z^4$ at infinity,
they define holomorphic sections of $L\subset W$
regular at $p^\pm_\infty$. They vanish at four points 
\qq
p'_\alpha=(V(z'_\alpha),\m z'_\alpha)\qquad\qquad{\rm or}\qquad
\qquad p''_\alpha=(-\m V(z''_\alpha),\m z''_\alpha)\,,
\qqq
respectively, where $z'_\alpha$ are the roots of $U$ 
and $z''_\alpha$ are those of $W$. Hence the degree 
of the line bundle $L$ is equal to 4. $H^0(L)$ has dimension
2 and is spanned by the two solutions (\ref{sol}).
\vskip 0.2cm

\subsection{Angle variables}

\noindent The knowledge of the bundle $L$ may be encoded in the
image $w$ of $L$ in the Jacobian $J^4(\CS)$ under the Abel map:
\qq
w\ =\ \sum\limits_{\alpha=1}^4
\int\limits_{p_0}^{p'_\alpha}
\Omega\ =\ 
\sum\limits_{\alpha=1}^4
\int\limits_{p_0}^{p''_\alpha}
\Omega
\label{angl}
\qqq
were $\Omega=(\Omega^b)$ is the vector of the abelian 
differentials (\ref{Omb}) on $\CS$ and $p_0$ is a fixed point 
of $\CS$. Under the infinitesimal
time evolution (\ref{dtn}) inducing the changes $\delta_nU$,
$\delta_nV$, $\delta_nW$, of the polynomials $U,V,W$,
the image of the Abel map changes by
\qq
\delta_nw^b\ =\ \sum\limits_{\alpha=1}^4{{z'_\alpha}^b\ 
\delta_n z'_\alpha\over V(z'_\alpha)}
\ =\ -\sum\limits_{\alpha=1}^4{{z''_\alpha}^b\  
\delta_n z''_\alpha\over V(z''_\alpha)}\s.
\qqq
The variations of the zeros of $U$ are
\qq
\delta_n z'_\alpha\,=\,-\,{\delta_n U(z'_\alpha)\over
U'(z'_\alpha)}
\qqq
and similarly for $\delta_nz''_\alpha$.
A direct calculation gives
\qq
\delta_n U(z)\,=\,\{\m h_n\m,\, U(z)\m\}\,\delta t_n\ 
=\ 4\, i\,\prod\limits_{m\not=n}(z_n-z_m)^{-1}\,{V(z_n)\m 
U(z)\m-\m U(z_n)\m V(z)\over z-z_n}\,\delta t_n\,,\hspace{0.7cm}
\qqq
see \cite{Mumf}, page 3.69. Hence
\qq
\delta_n w^b\ =\ 4\, i\prod\limits_{m\not=n}(z_n-z_m)^{-1}
\ U(z_n)\,\delta t_n\ \sum\limits_{\alpha=1}^4{{z'_\alpha}^b
\over(z'_\alpha-z_n)\,\m U'(z'_\alpha)}\,.
\qqq
The vanishing of the sum of residues of the meromorphic
form ${z^b\,\m dz\over (z-z_n)\, U(z)}$ implies that the last 
sum is equal to $-\m z_n^b\, U(z_n)^{-1}$ so that
\qq
\delta w^b\ =\ \{\m w^b\m,\, h_n\m\}\,\delta t_n
\ =\ {_4\over^i}\m\prod\limits_{m\not=n}(z_n-z_m)^{-1}
\,\m z_n^b\,\m\delta t_n
\qqq
which does not depend on the phase-space variables.
We infer that the Hamiltonians $h_n$ generate constant flows
on the complex torus $J^4(\CS)\cong\NC^3/\Lambda$ where 
$\Lambda$ is the lattice of periods of $\Omega$. 
Modulo a constant linear transformation, the coordinates 
$w^b$, $b=0,1,2,$ provide together with 3 of the Hamiltonians 
$h_n$ a Darboux coordinate system for $T^*\NP^3$. We have thus 
found the angle variables of the Hitchin system (as the angles
of $J^4(\CS)$).
\vskip 0.4cm

It should be stressed that the above approach based 
on the Lax matrix is simpler then the one obtained by
following the general procedure for the Hitchin systems. 
In particular, Eq.\s\s(\ref{det}) giving the spectral curve 
$\CC$ in the general approach is, as shown in \cite{GaTr},   
\qq
\xi^2\ =\ -\, 4\prod\limits_{n=1}^6(z-z_n)
\sum\limits_{n,m=1\atop n\not=m}^6{J_{nm}^{\s 2}
\over(z-z_n)(z-z_m)}
\qqq
to which one has to add the equation (\ref{curve}) 
of the original curve $\Sigma$ of genus 2. $\CC$ is 
of genus 5 and it is a ramified cover of both $\Sigma$ 
(by forgetting $\xi$) and $\CS$ (by setting 
$\sigma={i\over 2\sqrt{2}}\m\xi\m y$). While the 
general construction would give the angle variables
as those of a 3-dimensional Prym variety 
in the 5-dimensional Jacobian of degree $-2$ line bundles
on $\CC$, the Lax approach gave them as the angles
of the degree 4 Jacobian of $\CS$.
\vskip 0.2cm

\subsection{KZB connection}

\noindent The determinant bundle $\CL$ over the moduli space 
$\CN_{ss}\cong\NP\Theta_2$ of the holomorphic $SL_2$-bundles 
over the genus 2 curve coincides with the dual of the tautological 
bundle on $\NP\Theta_2$ so that $H^0(\CL^k)$ is the space 
of homogeneous polynomials $\Psi$ of degree $k$
on the space $\Theta_2$. The Lie algebra $so_{6}$ acts in
the space $\Theta_2$ by the first order differential 
operators, still denoted by $J_{nm}$, satisfying the commutation 
relations (\ref{soa1}), (\ref{soa2}) with the Poisson bracket replaced
by the commutator. In the $(p,q)$-language they are obtained
by replacing $p_n$'s in the expressions (\ref{Js}) for $J_{nm}$ by
$-\da_{x_n}$. The KZB connection for the case in question
has been work out in \cite{VGDJ}. It takes the form 
(up to a scalar 1-form)
\qq
&&\nabla^{\rm KZB}_{\delta\mu}\,\Psi\ =\ 
\sum\limits_{n=1}^6\delta z_n\,(\da_{z_n}
-{_1\over^{\kappa}}\m H_n)
\,\Psi\, ,\label{KZ21}\\
&&\nabla^{\rm KZB}_{\overline{\delta\mu}}\,\Psi\ =\ 
\sum\limits_n\overline{\delta z}_n\,\da_{\bar z_n}
\,\Psi
\label{KZ22}
\qqq
where 
\qq
H_n\ =\ -\m{_1\over^2}\sum\limits_{m\not=n}^6{J_{nm}^{\s 2}
\over z_n-z_m}
\qqq
so that ${1\over k} H_n$ is a quantization of ${2k\over
(2\pi i)^2}\m h_n$ obtained by the replacement $J_{nm}\mapsto 
{2\pi\over k\m i}\m J_{nm}$ in the classical expression 
for $h_n$. The quantum Hamiltonians $H_n,\ n=1,\dots,6$, 
are commuting second order differential operators on 
$\Theta_2\cong\NC^4$.

\nsection{Conclusions}

\noindent We have described above in explicit terms the
Hitchin integrable systems and the Knizhnik-
Zamolodchikov-Bernard connection
in the genus 0, 1 and 2 geometries, the last case only for $G=SL_2$
and with no punctures. The main original contribution of the paper
is the construction of the angle variables of the genus 2 system.  
It is a modification of a similar construction, based on the use 
of a Lax matrix, for the classical Neumann system. 
The diagonalization of the quantized Hamiltonians $H_n$
entering the genus 2 KZB connection for $G=SL_2$ as well
as the identification of the genus 2 Hitchin systems 
with punctures and for different groups remain open problems.


\end{document}